\documentclass[prb,superscriptaddress,twocolumn,showpacs,amsmath,aps]{revtex4}
\usepackage{epsfig,graphicx,hyperref}
\usepackage{bm}
\usepackage{subfigure}

\begin{document}

\title{Testing ``microscopic'' theories of glass-forming liquids}

\author{Ludovic Berthier} 
\affiliation{Laboratoire Charles Coulomb, CNRS-UMR 5221, 
Universit\'e Montpellier 2, Place Eug\`ene Bataillon, 
34095 Montpellier Cedex, France}

\author{Gilles Tarjus} 
\affiliation{LPTMC, CNRS-UMR 7600, Universit\'e Pierre et Marie Curie,
bo\^ite 121, 4 Pl. Jussieu, 75252 Paris Cedex 05, France}

\date{\today}

\begin{abstract}
We assess the validity of ``microscopic'' approaches of glass-forming liquids based on the sole knowledge of the static pair density correlations. To do so we apply them to a benchmark provided by two liquid models that share very similar static pair density correlation functions while displaying distinct
temperature evolutions of their relaxation times. We find that the approaches are unsuccessful in describing the difference in the dynamical behavior of the two models. Our study is not exhaustive, and we have not tested the effect of adding corrections by including for instance three-body density correlations. Yet, our results appear strong enough to challenge the claim that the slowdown of relaxation in glass-forming liquids, for which it is well established that the changes of the static structure factor with temperature are small, can be explained by ``microscopic''  approaches only requiring the static pair density correlations  as nontrivial input.
\end{abstract}

\pacs{05.10.-a, 05.20.Jj, 64.70.kj}

\maketitle

\section{Introduction}

Despite decades of intensive research, the connection between dynamics and structure in glass-forming liquids remains elusive. As a matter of fact, this is likely one main reason that explains the difficulty in producing a comprehensive theory of the glass transition\cite{tarjus11}. The structure of glass-forming liquids, as experimentally probed through the static structure factors or equivalently the pair correlation functions, does not appear to change much when temperature is decreased toward the glass transition while under the same condition, the relaxation time dramatically increases by orders of magnitudes. To incorporate this, rather central, observation, several working hypotheses have been proposed: (i) do away with the statics and assign the slowdown of relaxation to purely dynamical features such as emerging kinetic constraints\cite{palmer84,garrahan-chandler}, (ii) focus on structural quantities that go beyond the pair density correlations and represent higher-order and subtler correlations between particles\cite{bouchaud04,montanari06,cavagna07,steinhardt81,shintani06,sausset10,kurchan}, (iii) put forward mechanisms by which the small modifications of the static pair density correlations can be extremely amplified to produce spectacular dynamical changes. The theories based on (i) and (ii), such as those involving dynamical facilitation\cite{garrahan-chandler}, structural frustration\cite{nelson02,kivelson} or a random first-order transition\cite{wolynesRFOT,bouchaud04}, usually require some sort of phenomenological input, either because they involve some coarse-graining that introduces effective parameters only indirectly related to microscopic details or because they are intractactable in their full-blown generality and need approximate treatments. The paradigm of a type (iii) theory is the mode-coupling theory of the glass transition\cite{gotze} that predicts a freezing of the dynamics through a nonlinear feedback effect affecting the fluctuations of density.

The appeal of so-called ``microscopic'' theories is that they are able to make predictions from the knowledge of the molecular properties, among which the interaction potentials\cite{schilling03}. In the case of glass formation where it seems likely that some form of collective or cooperative behavior is at work and, correlatively, that some form of independence from the molecular details characterizes the phenomenology, it is not clear that this should necessarily be the best route to follow. In addition,  ``microscopic'' theories of the glass transition involve strong, uncontrolled approximations that are usually not justified at a microscopic level. Nonetheless, such theories are at least amenable to crisp tests on atomistic model systems.

In practice, ``microscopic'' approaches of the dynamics of glass-forming liquids are built on the knowledge of the static two-body density correlation functions (the static structure factors in Fourier space). The latter encode the microscopic information but represent only a partial description of the liquid structure. The issue we address here then boils down to the following question: how far can one go with the idea that the small observed changes in the static structure factors are sufficient to describe the slowdown of relaxation in glass-forming liquids?

The approaches we consider in this paper, with various degrees of exhaustivity and depth, are the mode-coupling theory (MCT), for which we shall mostly summarize the results obtained in our previous study\cite{berthier-tarjusMCT}, the ``microscopic'' implementations of the random first-order transition (RFOT) theory through density functional theory\cite{singh85,hall_DFT} or replica formalism\cite{mezard-parisi,zamponi10}, an approach that mix aspects of the MCT with thermally activated events\cite{schweizer}, and finally a proposed correlation between the dynamics of a glassformer and its static pair correlation functions via the two-body contribution to the excess entropy\cite{truskett}. To test these approaches, we have used as a benchmark the glass-forming systems that we have already extensively investigated by computer simulation\cite{berthier-tarjusPRL,berthier-tarjusJCP}: the Kob-Andersen binary Lennard-Jones model and its WCA (for Weeks-Chandler-Andersen\cite{WCA}) 
 reduction to truncated, purely repulsive potentials. These two models have been shown to share very similar static pair density correlation functions while displaying increasingly distinct evolution of their relaxation times as temperature decreases. A wide range of densities can moreover be considered, from low densities near the liquid-gas spinodal of the full binary Lennard-Jones mixture to high densities at which the difference between the two models progressively vanishes, through the conventional liquid/supercooled liquid range.

We find that the approaches based on the pair correlation functions, or equivalently on the static structure factors, are unsuccessful in describing the difference in the dynamical behavior of the two glass-forming models, the binary Lennard-Jones mixture and its WCA reduction. This is true even after applying for some theories a global rescaling of the predictions to better fit the simulation data for one model, say the binary Lennard-Jones one (in absolute terms, the predictions are indeed never very good): the difference in the slowdown of relaxation of the two models is still not correctly accounted for. As already mentioned, our study is not exhaustive, and we have not tested the effect of adding corrections to the description in terms of static pair correlations by including for instance three-body density correlations. In
addition, we have chosen to focus on simple glass-formers.
Yet, our results appear strong enough to challenge the claim that the viscous slowing down of glass-forming liquids, for which it is well established that the changes of the static structure factor with temperature are small, can be explained by ``microscopic''  approaches essentially  based on the knowledge of the static pair density correlations.

\section{Benchmark system}

The glass-forming systems that we consider are the three-dimensional Kob-Andersen binary Lennard-Jones mixture\cite{kob-andersen1994} (denoted LJ in the following) and its reduction to the purely repulsive part of the pair potentials proposed by Weeks, Chandler and Andersen\cite{WCA} (denoted WCA in the following). These are 80:20 mixtures of $A$:$B$ atoms with interatomic pair potentials
\begin{equation}
\label{eq_potentials}
\begin{aligned}
v_{\alpha \beta}(r) &= 4\epsilon_{\alpha \beta}\left[\left( \frac{\sigma_{\alpha \beta}}{r}\right)^{12}- \left( \frac{\sigma_{\alpha \beta}}{r}\right)^{6} + C_{\alpha \beta} \right] , \; {\rm for} \; r \leq r_{\alpha \beta}^c \\& =0, \; {\rm for} \; r \geq r_{\alpha \beta}^c,
\end{aligned}
\end{equation}
where $\alpha, \beta = A$  or $B$, $r_{\alpha \beta}^c$ is equal to the position of the minimum of $v_{\alpha \beta}(r)$ for the WCA model and to a conventional cutoff of $2.5 \sigma_{\alpha \beta}$ (merely introduced for practical reasons with no impact on the physical quantities) for the standard LJ model; $C_{\alpha \beta}$ is a constant that is fixed such that $v_{\alpha \beta}(r_{\alpha \beta}^c) =0$. The Molecular Dynamics simulations have been performed in the $NVE$ ensemble, after equilibration at the chosen temperature, with $N=1000$ particles, and we have studied a broad range of densities $\rho$ from $1.1$ to $1.8$
(a detailed description of the phase diagram is given in 
Ref.~\cite{berthier-tarjusJCP}).
Lengths, temperatures, and times are given in units of $\sigma_{AA}$, $\epsilon_{AA}/k_B$, and $(m \sigma_{AA}^2/48 \epsilon_{AA})^{1/2}$, respectively.

These systems form a benchmark for the kind of investigation that we want to carry out as their equilibrium pair structure is very close while their dynamics strongly diverge as temperature is lowered\cite{berthier-tarjusPRL,berthier-tarjusJCP}. Note that the two models should be considered at the same density (pressure is very sensitive to the presence or absence of attractive forces). The observables that we measure in the simulation are the partial  static structure factors $S_{\alpha \beta}(q)$ and pair correlation functions $g_{\alpha \beta}(r)$ as well as the self-intermediate scattering functions $F_s^{\alpha}(q,t)$ with $q$ near the position of the peak of the static structure factor at 
the most commonly studied liquid density
$\rho=1.2$ ($q\sigma_{AA}\simeq 7.2$); we extract the relaxation time from the latter, with the conventional choice $F_s^{\alpha}(q,t=\tau)=1/e$.

\section{Tested ``microscopic'' approaches}

The ``microscopic'' approaches of the dynamics of glass-forming liquids that we assess by comparing to simulation data on the above benchmark systems comprise first what can be taken as \textit{bona fide} theories, the mode-coupling theory (MCT)\cite{gotze} and the random first-order transition (RFOT) theory\cite{wolynesRFOT}. 

The MCT (for more detail, see our previous article\cite{berthier-tarjusMCT}) is based on a nonlinear differential equation for the time evolution of the two-point correlator of the density fluctuations. It can be derived for a Newtonian or for a Brownian dynamics, and, since the latter is somewhat simpler and leads to the same behavior at long times, this is the one that we have used. The time dependence of the intermediate scattering functions (two-point correlator of the density fluctuations) for the liquid mixture is then governed by
\begin{equation}
\label{eq_MCT}
\begin{aligned}
\frac{\partial}{\partial t} \mathbf{F}(q,t)=& -D_0\, q^2\, \mathbf{S}^{-1}(q) \mathbf{F}(q,t)\\& -\int_0^t dt' \mathbf{M}(q,t-t') \frac{\partial}{\partial t'} \mathbf{F}(q,t'),
\end{aligned}
\end{equation}
where $D_0$ is the diffusion coefficient of an isolated Brownian particle and $\mathbf{F}(q,t)$ is the matrix formed by the intermediate scattering functions $F_{\alpha \beta}(q,t)$ [such that $F_{\alpha \beta}(q,t=0)=S_{\alpha \beta}(q)$]. The physics is contained in the so-called memory kernel $\mathbf{M}(q,t)$ whose explicit form is derived through a series of approximations from the exact but intractable formal expression; it is a functional of the intermediate scattering functions $\mathbf{F}(q,t)$ (taken at the same time $t$) and, in the common implementation of the theory, of the static structure factors $\mathbf{S}(q)$:
\begin{equation}
\begin{aligned}
\label{eq_MCT_memory}
M_{\alpha \beta}(q,t)=&\sum_{\gamma \gamma' \delta \delta'} \int \frac{d^3k}{(2\pi)^3} \, \mathcal M^{\alpha \gamma \delta,\beta \gamma' \delta'}_{\mathbf q,\mathbf k}[ \mathbf{S}(q),\mathbf{S}(\vert \mathbf q - \mathbf k \vert)]\\& \times F_{\gamma \gamma'}(\vert \mathbf q - \mathbf k \vert,t)\, F_{\delta \delta'}(k,t),
\end{aligned}
\end{equation}
where the expression of $\mathcal M^{\alpha \gamma \delta,\beta \gamma' \delta'}_{\mathbf q,\mathbf k}$ is given in Ref.~[\onlinecite{flenner05}]. (Note that in principle this expression involves the static triplet correlation function but the latter is almost always considered in a factorized approximation.) In this formulation, the only nontrivial input is therefore the partial static structure factors $S_{\alpha \beta}(q)$.  An equation can also be derived for the intermediate scattering functions $F_s^{\alpha}(q,t)$\cite{nagele99}. However, it is known that for wave-vectors corresponding to typical interatomic distances the temperature dependences of the relaxation times associated with the collective and the self-intermediate scattering functions are similar and, moreover, that the predicted critical temperatures $T_c$ at which the dynamics freezes are identical\cite{gotze}.

The second type of theory that we shall test are microscopic implementations of the  random first-order transition (RFOT) theory. The latter builds on an analogy with the behavior of mean-field spin glasses without reflection symmetry\cite{kirkpatrick-wolynes,kirkpatrick-thirumalai}. It provides a scenario for the slowdown of relaxation in glass-forming liquids that focuses on the free-energy landscape, which is postulated to be characterized by an exponentially large number of ``metastable'' states and the associated configurational entropy (or complexity), and on an entropy-driven activated relaxation, which leads to a heterogeneous ``mosaic'' liquid state\cite{wolynesRFOT}. Microscopic implementations of the RFOT theory can be realized at a mean-field level where free-energy ``metastable'' states have an infinite life-time and are therefore properly defined. They can be formulated either in terms of a density functional approach (DFT)\cite{singh85,hall_DFT,dasgupta_DFT} or within a replica formalism\cite{monasson95,mezard-parisi,franz-parisi,zamponi10}. In these mean-field-like approaches, one does not have a direct access to the dynamics of the system but indirect information is provided through the configurational entropy and the two critical temperatures, $T_d$ at which the complexity jumps to a finite value and $T_K$ at which the complexity vanishes.

In the DFT treatment, one looks for the density profiles that minimize some appropriately derived (mean-field) density functional. The main steps are the formulation of the density functional itself and the \textit{a priori} characterization of the structure of the metastable states in terms of trial (amorphous or ``aperiodic''\cite{wolynes_aperiodic}) density profiles. The commonly considered Helmoltz free-energy functional is the Ramakrishanan-Yussouff\cite{rama-yussouf79} approximation which involves a Taylor series expansion (truncated after the quadratic term) of the excess free energy around the liquid phase with mean density $\rho_0$.  For a one-component liquid, it is expressed as
\begin{equation}
\label{eq_DFT}
\begin{aligned}
&\mathcal F[\rho]= \int d\mathbf r\, \rho(\mathbf r) \left (\ln[\Lambda^3\, \rho(\mathbf r)] - 1\right ) + \mathcal F^{ex}(\rho_0)\\&-\frac{1}{2}\int d\mathbf r_1 \int d\mathbf r_2\, c(\vert \mathbf r_1-\mathbf r_2\vert;\rho_0)[\rho(\mathbf r_1)-\rho_0][\rho(\mathbf r_2)-\rho_0],
\end{aligned}
\end{equation}
where the first term of the right-hand side is the ideal-gas functional, with $\Lambda$ the thermal wavelength, and $\mathcal F^{ex}(\rho_0)$ is the excess free energy of a homogeneous liquid of density $\rho_0$; $c(r;\rho_0)$ is the direct correlation function of the homogeneous liquid and is related to the pair correlation function $g(r;\rho_0)$ through the Ornstein-Zernike integral equation\cite{hansen-macdo}. The density profiles $\rho(\mathbf r)$ of the metastable states are generally described through a sum of Gaussian functions centered about a given amorphous (or aperiodic) lattice, with the width of the Gaussians characterized by a variational localization parameter on which the minimization of the free energy is then performed. In practice, the amorphous lattice is chosen from a random close packing of hard spheres with some effective diameter. Additional approximations can also be introduced in order to further simplify the computations\cite{singh85,hall_DFT,dasgupta_DFT}.

An alternative approach uses the replica formalism, which allows one to directly study the statistical properties of the metastable states and the configurational entropy. To this end, one introduces $m$ replicas of the original liquid system and, if one is only interested in the liquid above the putative RFOT to an ideal-glass phase, one considers the difference between the free energy in the limit $m \rightarrow 1^+$ (after having taken the thermodynamic limit) and the homogeneous liquid value obtained for $m=1$\cite{monasson95,mezard-parisi}. At low enough temperature, one expects the atoms in the $m$ copies to cluster around around common positions in space, thereby forming a ``molecular bound state'' and one can use a ``small cage expansion'' around typical liquid configurations in the spirit of the Einstein description of a crystal (in the DFT context, this would correspond to the self-consistent phonon approximation\cite{stoessel84}). Truncated to quadratic (harmonic) order, this approximation leads to the following free energy per particle
\begin{equation}
\label{eq_replicas}
\begin{aligned}
\phi(m,A;T)= &\frac{3(1-m)}{2m}k_B T\left [\ln(2\pi A) + 1\right ]- m A\,C_{liq}(T/m) \\&-\frac{3}{2m}k_B T\ln (m) +\frac{1}{m}\phi_{liq}(T/m),
\end{aligned}
\end{equation}
where $A$ is the cage size, whereas $C_{liq}(T)$ and $\phi_{liq}(T)$ are the expectation value of the Laplacian of the potential and the free-energy density for the equilibrium liquid (\textit{i.e.} with $m=1$) at temperature $T$. The free energy in the above equation should be minimized with respect to $A$ (in the glass phase which we shall not consider here one also has to minimize with respect to $m$). The complexity (or configurational entropy per particle) in the equilibrium liquid is given by
\begin{equation}
\label{eq_replicas_complexity}
\begin{aligned}
\Sigma(T)= \frac{m^2}{k_B T}\frac{\partial}{\partial m}\bigg[\frac{\phi(m,A^*(m;T);T)}{m}\bigg ]\bigg\vert_{m\rightarrow 1},
\end{aligned}
\end{equation}
where $A^*(m;T)$ is the cage size at the minimum of Eq.~(\ref{eq_replicas}).
Because we only study the the temperature range above the ideal glass 
transition where $m=1$, an expression of the complexity in this range
is easily obtained by expanding Eq.~(\ref{eq_replicas}) around $m=1$, 
such that the definition in Eq.~(\ref{eq_replicas_complexity}) yields  
\begin{equation}
\Sigma(T) = S_{liq}(T) - \frac{3}{2} \left[ 1+ \ln 
\left(  \frac{3T}{2 \langle \Delta V(r) \rangle(T)} 
\right) \right], 
\end{equation}
where $\langle \Delta V(r) \rangle = 2 \pi \rho \int_0^\infty 
dr \Delta V(r) g(r)$ and $\Delta$ is the Laplacian.  
This expression shows that the complexity is obtained by substracting 
from the total entropy a ``vibrational'' term requiring the sole 
knowledge of the pair correlation function $g(r)$. 

Whereas the two approaches give in principle the same information, namely the complexity, the DFT is more demanding as it requires a choice of trial configurations to describe the density profiles corresponding to the glassy metastable states. Except for some comments on the DFT, we shall therefore illustrate the ability of such calculations to reproduce the differing behavior of the WCA and LJ models by studying the complexity and $T_K$ via the replica formalism and the small cage expansion. We only seek to give an idea of what the approach can or cannot do (with the commonly used approximations) and we therefore further simplify the computation by obtaining the $g(r)$ for the WCA and LJ models via the HNC integral equation for a one-component system.
Of course, the choice of the HNC integral equation may have
some quantitative impact on the results.

We next consider an approach that mixes microscopic input and phenomenological considerations. Schweizer and coworkers\cite{schweizer} have proposed a somewhat heuristic extension of the MCT that incorporates activated mechanisms allowing ergodicity to be restored at temperatures below the $T_c$ of MCT. It focuses on single-particle dynamics and boils down to a stochastic nonlinear Langevin equation for the displacement of a particle $r(t)$. The central quantity is the ``nonequilibrium free-energy functional'' which gives rise to the effective force exerted by the surrounding on a tagged particle. For a one-component liquid, this functional reads
\begin{equation}
\label{eq_Schweizer}
\begin{aligned}
&F_{eff}(r)/(k_B T)= -3 \ln(r) \\&-\int \frac{d^3q}{(2\pi)^3}\, \frac{(S(q)-1)^2}{\rho(S(q)+1)} \exp\left[-\frac{q^2 r^2}{6}(1+S^{-1}(q))  \right]
\end{aligned}
\end{equation}
where $S(q)$ is the static structure factor. In the viscous liquid regime, the competition between the two terms in Eq.~(\ref{eq_Schweizer}) produces a minimum at small $r$ and a maximum at a larger value. The relaxation time is given by the hopping time to escape from localization in the minimum and 
is expressed following the standard Kramers theory as
\begin{equation}
\label{eq_Schweizer_tau}
\tau(T) = \tau_0 \exp \left [\frac{F_b(T)}{k_B T}\right ],
\end{equation}
where $F_b$ is the height of the barrier between the maximum and the minimum of $F_{eff}(r)$ and the prefactor $\tau_0$ includes,
in addition to the short-time friction constant,
information about the curvature around the minimum and the maximum.
Since it
has been shown to be weakly dependent on temperature and density, we take it
as a constant.

Finally, we also assess a correlation between dynamics and static pair density correlation that has recently been put forward, mostly on an empirical basis. Building on Rosenfeld's work\cite{rosenfeld}, Truskett and coworkers\cite{truskett,mittal06} have proposed a direct connection between thermodynamics and dynamics in liquids. In their picture, the transport and relaxation properties of a liquid are determined by the thermodynamic excess entropy (the entropy in excess to that of the ideal gas, not to be confused with the ``configurational entropy'' in excess to that of the crystal). For simple atomic liquids in which only translational degrees of freedom are relevant, Truskett and coworkers went further to replace the excess entropy by its two-body contribution $s_2$ and therefore proposed a functional relationship between the diffusivity or the relaxation time and the static pair correlation function $g(r)$\cite{truskett,chopra10}. For a one-component system, the two-body excess entropy is indeed defined as
\begin{equation}
\label{eq_Truskett1_s2}
\begin{aligned}
- s_2(T)/k_B= \frac{\rho}{2} \int d^3r\, \bigg[g(r) \ln [g(r)] - g(r) +1\bigg],
\end{aligned}
\end{equation}
which for a binary mixture is generalized to
\begin{equation}
\label{eq_Truskett2_s2}
\begin{aligned}
&- s_2(T)/k_B=\\&\frac{\rho}{2}\sum_{\alpha \beta}x_{\alpha}x_{\beta} \int d^3r\, \bigg[g_{\alpha \beta}(r) \ln [g_{\alpha \beta}(r)] - g_{\alpha \beta}(r) +1\bigg],
\end{aligned}
\end{equation}
where $x_{\alpha}$ is the concentration of species $\alpha$.
We only focus here on the application of these ideas to the viscous liquid regime.

The authors\cite{truskett} moreover suggested that a good empirical expression to link transport and relaxation properties to two-body excess entropy is of the form
 \begin{equation}
\label{eq_Truskett_exp}
\begin{aligned}
\tau(T) \propto \exp\left [-K s_2(T)\right],
\end{aligned}
\end{equation}
with $K$ an adjustable parameter. (It should be stressed that the above equation is quite different than the much used Adam-Gibbs formula\cite{adam-gibbs65}, as the entropy comes in the numerator of the term in the exponential whereas the configurational entropy comes in the denominator in the Adam-Gibbs expression.) Note that this connection between excess entropy and relaxation has already been shown to fail in the case of associated liquids (square-well fluids\cite{trus1} and water models\cite{trus2}) and for confined liquids\cite{trus3} in the supercooled regime.

All of the above descriptions of the dynamics of glass-forming liquids, at the level of approximation which is that commonly used but could in principle be improved (see below),  involve only the static pair correlation functions as nontrivial input, as shown by Eqs.~(\ref{eq_MCT}, \ref{eq_MCT_memory}), (\ref{eq_replicas}, \ref{eq_replicas_complexity}), (\ref{eq_Schweizer}, \ref{eq_Schweizer_tau}), and (\ref{eq_Truskett1_s2}-\ref{eq_Truskett_exp}). In the following, we confront these approaches to the benchmark systems described above.

\section{Results}

A difficulty in assessing the ability of the microscopic approaches based on the knowledge of the static pair density correlations to reproduce the differing behavior of the WCA and LJ glass-formers is that, when tested in simulation studies of glass-forming liquid models, these descriptions anyhow never provide a very accurate account of the dynamical data. This is somewhat expected as the approximations involved in the theoretical formulations are quite uncontrolled and serious, while the phenomenology to be described is quite rich and complex.
In approaches such as
the MCT and the RFOT theory, a way to
get around this is to implement some rescaling procedure to make predictions and simulation data on one type of observable fare as well as possible and use this rescaling to compare the predictions for other observables or phenomena. This is for instance what is usually done in test studies of the MCT: the critical temperature $T_c$ is rescaled in order to better match the simulation data on the relaxation time and comparisons are then made at relative distance from this temperature\cite{kob_MCT}. In the following, we shall allow for a possible rescaling of the theoretical predictions for one model, the LJ one, and use the \textit{same} rescaling parameter for the WCA model. Doing otherwise by allowing for different rescaling parameters would completely side-step the issue of comparing these two systems, which have very close static structure factors and very different dynamics at the same ($\rho$, $T$) state point. 

We first consider the results for the MCT. A detailed account has been given in Ref.~[\onlinecite{berthier-tarjusMCT}] and we illustrate here our findings for the sake of completeness. We have solved numerically Eqs.~(\ref{eq_MCT}, \ref{eq_MCT_memory}) by using as input the partial static structure factors obtained in our simulation of the WCA and LJ models. From the solution for the intermediate scattering functions $F_{\alpha \beta}(q,t)$, we have extracted a relaxation time $\tau$ (defined from $F_{\alpha \alpha}(q,t=\tau)/S_{\alpha \alpha}(q)=1/e$ for the majority species $\alpha=A$ and for $q\sigma_{AA}\simeq 7.2$) and we have determined the critical temperature $T_c$.

\begin{figure}
\psfig{file=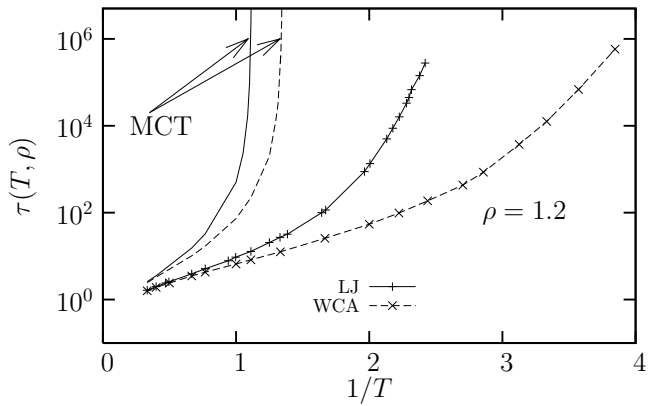,width=8.5cm}
\caption{\label{figure_MCT_Arrhenius}
Comparison of the relaxation times obtained by MCT (lines) and simulations 
(symbols) for the LJ (full lines) and the WCA (dashed lines) 
models at a density $\rho=1.2$.}
\end{figure}

\begin{figure}
\psfig{file=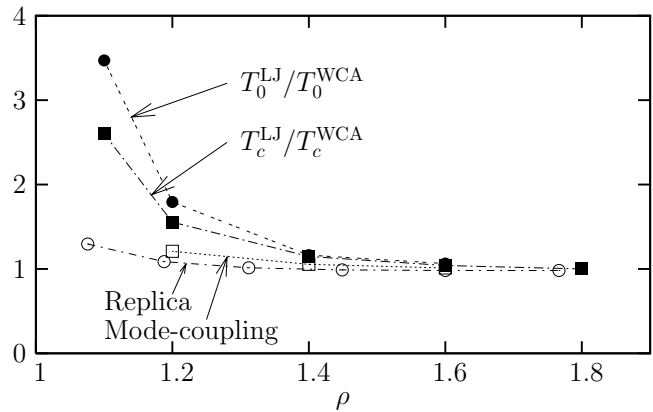,width=8.5cm}
\caption{\label{figure_tau_ratios}
Comparison between the predictions of the MCT and of the replica-RFOT approach and the simulation results. Squares: ratio of the critical temperatures $T_c$ for the LJ and WCA models as predicted by MCT (open symbols) and estimated from fits to the simulation data (closed symbols). 
Circles: ratio $T_K^{\rm LJ}/T_K^{\rm WCA}$ predicted in the replica approach (open symbols) and of the ratio $T_0^{\rm LJ}/T_0^{\rm WCA}$ estimated from
the Vogel-Fulcher-Tammann fit to the simulation data
for the relaxation time (closed symbols). }
\end{figure}

To begin with, we give in Fig.~\ref{figure_MCT_Arrhenius} an example of the difficulty mentioned above: on an Arrhenius plot of the relaxation time at $\rho=1.2$, the MCT predictions are quantitatively inaccurate even for the LJ model and one must rescale the temperature to obtain a more acceptable description. From the bare data, one can clearly see that the difference between the slowdown of relaxation of the WCA and the LJ models is considerably underestimated by the theory. To disentangle this from the overall quantitative inaccuracy, we plot in Fig.~\ref{figure_tau_ratios} the ratio of the critical temperatures $T_c$ of the LJ and the WCA models (which automatically accounts for a rescaling of temperature that is the same for the two models) as predicted by the theory (see preceding section) and obtained from a fit to the simulation data in the range of temperature where this fit is possible~\cite{berthier-tarjusMCT}: the trends as function of density are similar but the theory is unable to capture the wide difference in the dynamics of the two models at typical liquid density ($\rho \lesssim 1.4$).

\begin{figure}[b]
\psfig{file=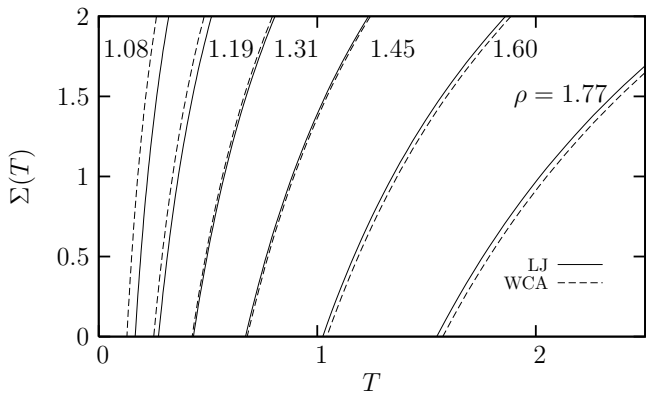,width=8.5cm}
\caption{\label{figure_replicas_complexity}
Temperature dependence of the predicted complexity (from the replica 
formalism and the small-cage expansion)
of the LJ (full lines) and WCA (dashed lines) one-component models for several densities. A vanishing complexity defines the ideal 
glass transition temperatures, $\Sigma(T_K)=0$.}
\end{figure}

For the replica (mean-field) approach of the RFOT theory, we have solved the HNC integral equation for the WCA and LJ models and used the resulting pair correlation function as input in Eqs.~(\ref{eq_replicas}, \ref{eq_replicas_complexity}) to compute the complexity as a function of temperature and density as well as the associated $T_K(\rho)$ at which the complexity goes to zero. The results showing the complexity as a function of temperature for several densities are given in Fig.~\ref{figure_replicas_complexity}. One observes that the curves for the WCA and LJ models are always very close (the  difference is slightly bigger when one lowers the density to values close to the spinodal of the LJ liquid). To give some elements of comparison between theory and simulation, we have fitted the temperature dependence of the relaxation data of the WCA and LJ liquids with a Vogel-Fulcher-Tammann expression, $\tau = \tau_0 \exp[B T_0/(T-T_0)]$. As often done, we take $T_0$ as an estimate for the $T_K$ of the ``entropy crisis''. We are not interested here in the absolute values but in the ratio of the temperatures for the WCA system and for the LJ one. We plot in Fig.~\ref{figure_tau_ratios} the evolution with density of the theoretical prediction for the ratio $T_K^{\rm LJ}/T_K^{\rm WCA}$ and of the simulation result for the ratio $T_0^{\rm LJ}/T_0^{\rm WCA}$. Without any doubt, the theory completely misses the widening gap between the two models as density decreases.

Concerning the DFT approach to the RFOT theory, which we have not examined in detail, we just make the following remark. If, as done \textit{e.g.} in Ref.~[\onlinecite{hall_DFT}], the approximation to the DFT approach relies on a WCA-like separation of the pair potentials with (i) the equilibrium structure (\textit{i.e.}, both the pair correlation function $g(r)$ and the reference aperiodic lattice) determined by the truncated repulsive component and (ii) the attractive interaction treated in a mean-field-like fashion, then, essentially by construction, it cannot describe the difference in behavior between the two models, with (LJ) and without (WCA) attractive tails. The fact that Hall and Wolynes\cite{hall_DFT} use a somewhat different choice of cutoff than the conventional WCA one does not change this conclusion. 
The modified KRR truncation~\cite{KRR} that they
consider is equivalent to the WCA one for typical liquid densities (here
$\rho \sim 1.2$); as implied from Fig. 5 of 
Ref.~\cite{hall_DFT}, the predictions fare better at lower
densities (corresponding here to $\rho=1.1$) but they do worse at larger
densities where they even fail to recover the merging with the LJ data
found for the WCA truncation.

\begin{figure}
\psfig{file=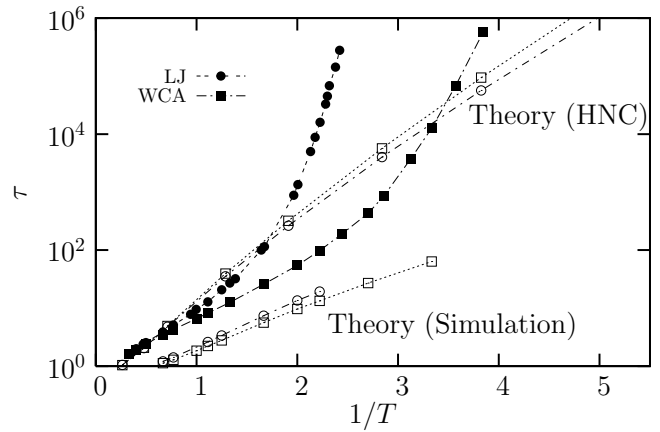,width=8.5cm}
\caption{\label{figure_schweizer1}
Arrhenius plot of the relaxation time for the LJ and WCA models as predicted from the Schweizer-Saltzman approach (open symbols) and obtained from simulation (filled symbols). For the theory, we show both the results of the effective one-component approximation with static structure data from simulation as input (lower curves) and those for the one-component models with pair correlation function obtained from the HNC integral equation 
(upper curves). Note that both theoretical predictions evolve with 
an unphysical sub-Arrhenius manner with temperature.}
\end{figure}

To test the Schweizer-Saltzman approach for binary liquid mixtures, one should in principle consider a two-dimensional  extension of Eq.~(\ref{eq_Schweizer}) which takes into account the displacements of particles of both species. However, this is numerically very demanding. We have taken instead a shortcut that uses an effective one-component description focusing on the majority component $A$ of the $80$:$20$ binary mixture. To do so 
we have implemented the procedure developed in Ref.~\cite{viehman-schweizer} 
in a different context of a mixture of attractive and repulsive spheres.
This approximation amounts to replacing $\rho$ and $S(q)$ in the one-component
expression in Eq.~(\ref{eq_Schweizer}) by $\rho_A$ and $S_{AA}(q)$, respectively.
Thus, we can directly feed the theory with the structure factors measured 
in the numerical simulations, much as we did to study 
mode-coupling  theory. Additionally, we have solved the HNC integral
equation for the one-component LJ and WCA potentials at the density 
$\rho=1.2$, which we then
directly use in Eq.~(\ref{eq_Schweizer}).

We plot in Fig.~\ref{figure_schweizer1} the predictions for the relaxation time as a function of $1/T$ at the density $\rho=1.2$, together with the simulation data. The striking observation is that the theoretical predictions fail 
completely on two key aspects. First they fail to reproduce the super-Arrhenius behavior with its marked curvature. Second, they also fail to predict any significant difference between the WCA and the LJ models. Puzzled by this major breakdown of the approach, we have checked our implementation in the case of a 
hard-sphere system
already studied by Schweizer and Saltzman\cite{schweizer}. The results are presented in more detail in the Appendix and confirm the trends observed for the LJ system: the slowdown of relaxation as one cools or compresses a glass-forming system is strongly underestimated by the Schweizer-Saltzman approach. It is true that the effective one-component treatment introduces an additional approximation, but considering that we apply it to the majority component of the LJ and WCA mixtures which comprises $80 \%$ of the atoms, it seems more likely that the observed failure is intrinsic to the theoretical approach.

\begin{figure}
\psfig{file=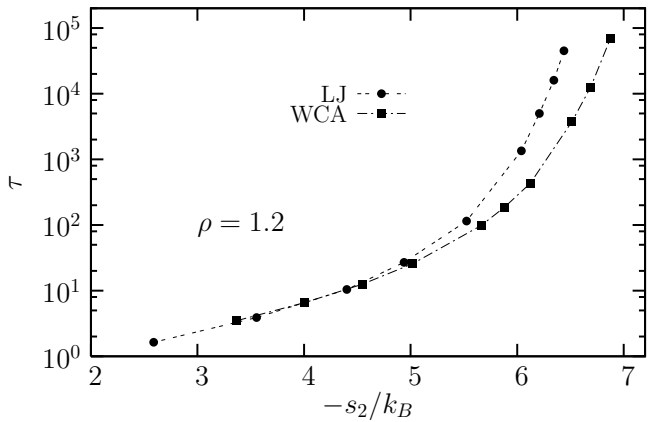,width=8.5cm}
\caption{\label{figure_truskett}
Relaxation time $\tau$ versus two-body excess entropy $-s_2/k_B$ for the LJ (circles) and WCA (squares) liquids at density $\rho=1.2$ (from simulation data).}
\end{figure}

Finally, we come to the correlation between the relaxation time and the two-body part of the excess entropy. We have computed the latter from the simulation data for the partial pair correlation functions according to Eq.~(\ref{eq_Truskett2_s2}). The plot of $\tau$ versus $-s_2/k_B$ is shown in Fig.~\ref{figure_truskett} for the typical liquid density $\rho=1.2$. The data for the WCA and LJ models collapse at high temperature (smallest values of $-s_2/k_B$) but markedly deviate as $T$ decreases (and $-s_2/k_B$ increases), clearly showing the absence of master-curve. We note that the growing gap between the WCA and LJ data is somewhat reduced when plotted versus $s_2$ in place of $1/T$ but the results nonetheless demonstrate that the two-body excess entropy does not uniquely determine the dynamics.
Similar conclusions have been previously reached concerning supercooled associated liquids~\cite{trus1,trus2} and liquids under confinement\cite{trus3}.

\begin{figure}
\psfig{file=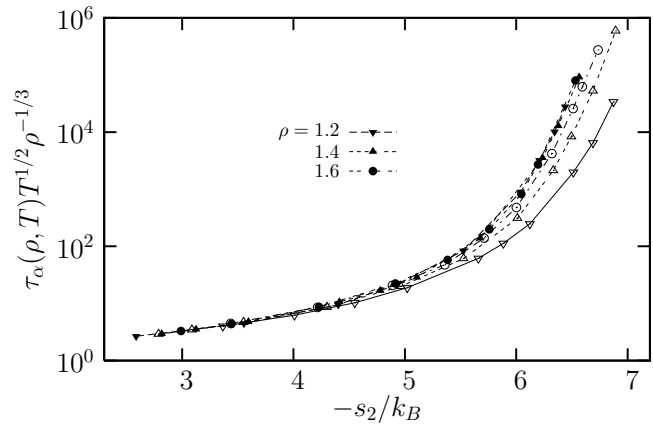,width=8.5cm}
\caption{\label{figure_truskett_scaling}
Relaxation time $\tau$ (normalized by a high-$T$ value $\tau_{\infty} \propto \rho^{1/3}T^{-1/2}$) versus two-body excess entropy $-s_2/k_B$ for the LJ and WCA liquids at several densities (from simulation data). Note the good collapse of data for the LJ model (filled symbols) and its absence for the WCA one (open symbols).}
\end{figure}

We have also studied the evolution with density of the proposed correlation. It is expected, and indeed found, that the curves $\tau$ versus $-s_2/k_B$ for the WCA and LJ models converge as density becomes high enough. We have plotted in Fig.~\ref{figure_truskett_scaling} the relaxation time  (normalized by a high-temperature value $\tau_{\infty}$ which has a small residual dependence on density and temperature) versus $-s_2/k_B$ for the two models. Besides showing the above mentioned convergence at high density, one observes that the data for the LJ liquid at all densities  collapse quite well on a master-curve, whereas a markedly different behavior is seen for the WCA liquid. This illustrates that the two-body excess entropy does not capture the difference in the density evolution of the slowdown of relaxation, which can be rescaled in the case of the LJ model but not for the WCA one\cite{berthier-tarjusPRL,berthier-tarjusJCP}. This can be easily understood. Roughly speaking, and as far as fluctuations are concerned, the LJ liquid behaves as a soft-sphere model\cite{dyre} with power-law repulsive interactions characterized by an effective exponent $\gamma$. The relaxation time is then a function of the scaling variable $\rho^{\gamma}/T$ only (as already shown\cite{dyre,coslovich-roland,berthier-tarjusPRL}), but this should also be true for the two-body excess entropy: one then expects a collapse of the $\tau$ versus $-s_2/k_B$ data, as indeed observed in Fig.~\ref{figure_truskett_scaling} (this indicates that the adjustable parameter $K$ in Eq.~(\ref{eq_Truskett_exp}) is essentially independent of density). However, as the pair correlation functions of the WCA and LJ models are  very close, the two-body excess entropy of the WCA liquid has also an approximate scaling in $\rho^{\gamma}/T$. It is then obvious that the strong violation of density scaling which is observed in the slowdown of relaxation for this model\cite{berthier-tarjusPRL,berthier-tarjusJCP} cannot be accounted for by the two-body excess entropy
. This is what we see in Fig.~\ref{figure_truskett_scaling}.

\section{Discussion}

The so-called ``microscopic'' approaches of the dynamics in glass-forming liquids that rely on the sole knowledge of the static pair density correlations fail when tested against the benchmark provided by two glass-forming liquid models with very similar static pair structure and strongly different dynamics. This is so even if one adjustable rescaling parameter, the same for the two models, is introduced to improve the overall quantitative accuracy with respect to simulation data. Of course, by introducing more adjustable parameters, one would end up with a better description but this would obviously be at the expense of the ``microscopic'' character of the approaches. Our demonstration is not flawless, as we have introduced in some cases an additional assumption in the form of an effective one-component description of the binary mixture. However, we believe it is strong enough to seriously question the use of theoretical or empirical descriptions of the dynamics of glass-forming liquids that are based on the static pair density correlations only. This finding echoes with extreme theoretical models  
which have been devised where the slowdown of relaxation 
takes place in the complete absence of 
pair correlations among particles~\cite{barrat,schilling,frenkel}. 

Some of the approaches studied here could in principle incorporate more information on the structure than just the pair correlation function. This is not what is done in their common implementations, and this would certainly increase the complexity of the computations while reducing the simplicity of the message. Although we feel that merely introducing corrections due to the triplet correlation functions in the MCT or the DFT approach of the RFOT theory would not be sufficient, we cannot exclude that considering more involved approximations that account for higher-order structural correlations would lead to a better description of our benchmark systems\cite{daniele}. In the DFT approach of the RFOT theory, one could for instance envisage considering trial amorphous configurations that already account for the subtle structural differences between the WCA and the LJ liquid; or, in the sought correlation between dynamics and excess entropy, one could consider the exact excess entropy as in Ref.~[\onlinecite{mittal06}] in place of its two-body contribution. It remains to be seen whether this improves the predictions, and we hope that the present work will stimulate further investigations along these lines.

We would like to stress again that the above conclusions apply to glass-forming \textit{liquids}, for which it is well established that the dramatic slowdown of relaxation comes with only minor changes in the equilibrium pair density correlations. For fluids able to be taken at low temperature and low density, such as systems interacting with truncated repulsive potentials, possibly supplemented by short-ranged attractive interactions, a wide thermodynamic range can be covered within which the static pair correlation functions may show marked and nontrivial evolution\cite{att_colloids,harm_spheres}. The ability of theories based on the static pair structure to grasp some qualitative and quantitative trends is then less challengeable in this case.

Finally, we note that the present study has obviously nothing to say about theories of glass formation that rely, at least in principle, on more than the static pair density correlations, involving either higher-order static correlations or kinetic constraints (see Introduction). Such theories however are not microscopic in the sense that phenomenological input is necessary, either because the full blown theory is intractable (at present) or because some pieces connecting to the microscopic details are still missing. A crisp test of the type proposed here is therefore harder to envisage.

\appendix*
\section{Schweizer-Saltzman approach for hard spheres}

In this appendix, we check our implementation of the Schweizer-Saltzman approach for the binary LJ and WCA liquids by applying it to the hard-sphere model already extensively studied by these authors\cite{schweizer}. 

We obtain the static pair structure using three methods: we solve 
both the HNC and PY integral equations for a monodisperse system 
of hard spheres to get $S(q)$, or we follow the same procedure
as in the main text and measure $S_{AA}(q)$ for the majority
component of a binary mixture of hard spheres to use the same 
effective one-component approximation. We then
inject this structural information into
Eq.~(\ref{eq_Schweizer}) to obtain a prediction for the 
relaxation time which we represent either as a function 
of the volume fraction $\phi$, or as a function of the reduced 
pressure (or compressibility factor), $Z =  P /(k_B T \rho)$. 
The latter is  evaluated from the Carnahan-Stirling
equation of state, which is quite accurate even in the high-$\phi$
regime that we wish to discuss~\cite{EOS}. 

\begin{figure}
\psfig{file=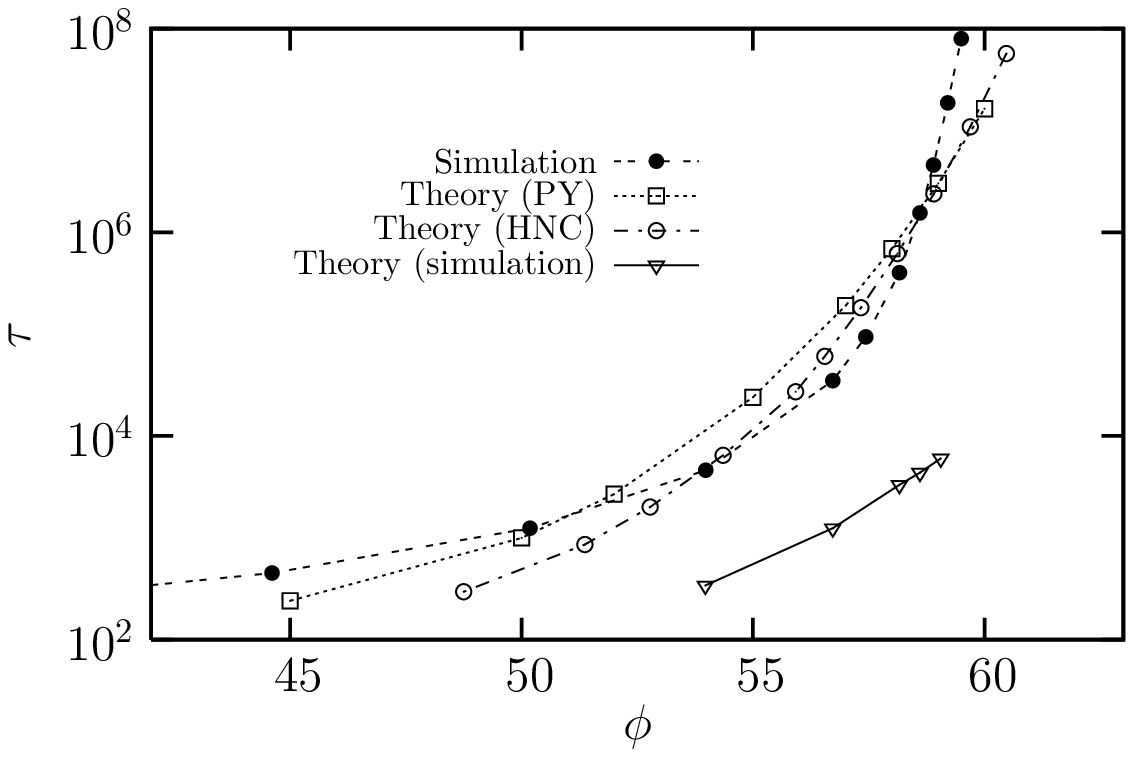,width=8.5cm}
\psfig{file=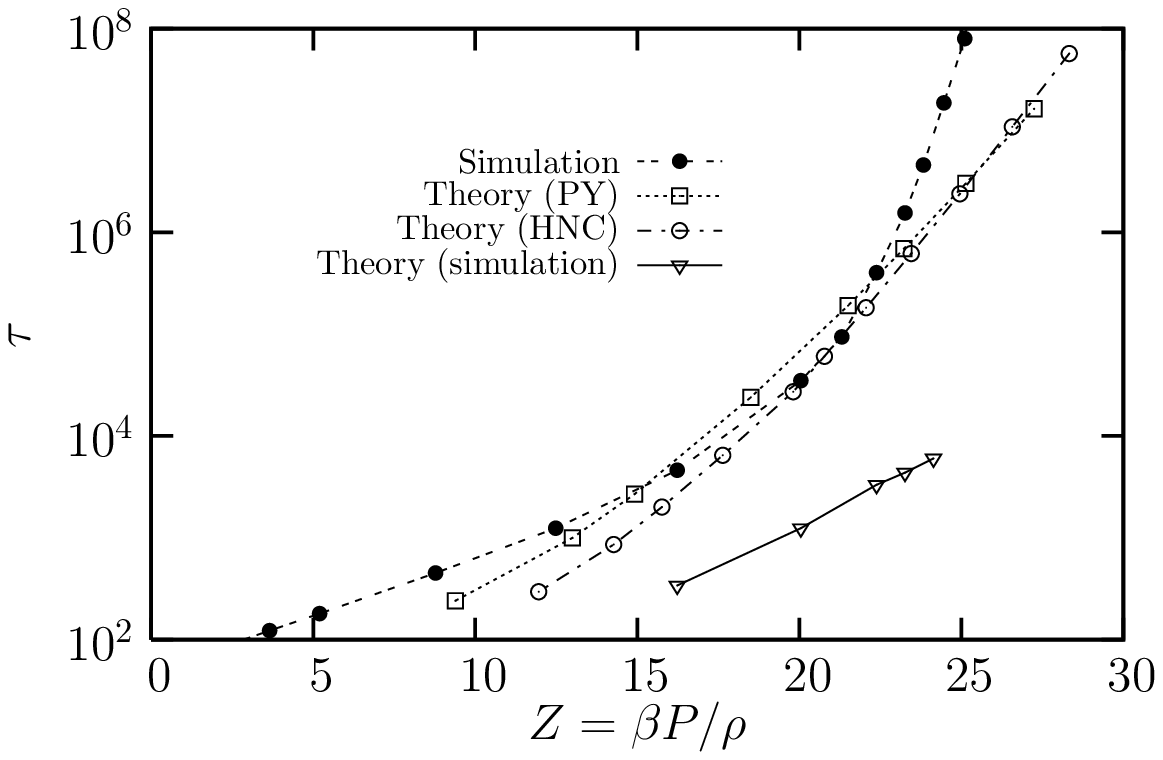,width=8.5cm}
\caption{\label{figure_schweizer_HS1}
Top: Logarithmic plot of the relaxation time $\tau$
versus packing fraction $\phi$ for the hard-sphere systems.
Predictions from the Schweizer-Saltzman approach with
the $g(r)$ obtained from the PY (squares) and the HNC (circles) closures
for the monodisperse system, as well as from simulation of the binary
mixture using the effective one-component treatment (triangles).
We also display the simulation data (filled symbols)
from Ref.~[\onlinecite{berthier_HS}] for the binary mixture.
Bottom: same data plotted as a function of
$Z=P/(k_B T\rho)$.}
\end{figure}

The results of the two theoretical approaches using HNC and PY are displayed in 
Fig.~\ref{figure_schweizer_HS1}, and they fully match 
published results\cite{schweizer}, which shows that we have correctly 
implemented the Schweizer-Saltzman approach. It should be noted that
when plotted in the representation $\log(\tau)$ versus $Z$, these data 
appear almost linear, which seems to suggest that 
in the $\phi$ regime relevant to most simulations and experiments, 
the Schweizer-Saltzman approach seems to yield results similar to the 
free volume approach $\tau \sim \exp(Z)$, and that the asymptotic
regime predicted by Schweizer\cite{schweizerjamming}, 
$\tau \sim \exp(Z^2)$, is not entered.

It is not possible to obtain numerical results for a monodisperse 
system of hard spheres to assess these theoretical predictions 
directly because
the system crystallizes too easily. However, it is well-known that 
slightly polydisperse systems and binary mixtures can easily be compressed
toward the glass transition with no sign of crystallisation 
and very little effect of polydispersity on the actual
location of the glass ``transition''\cite{HSpoly1,HSpoly2}.
Thus, we use numerical results\cite{berthier_HS,berthier_witten} 
obtained for 
a $50$:$50$ mixture of hard spheres with diameter ratio $1.4$, and 
present them along with the theoretical results in 
Fig.~\ref{figure_schweizer_HS1}. This comparison appears reasonable as Schweizer and Saltzman themselves compare 
their predictions with results obtained from simulations 
and experiments performed on polydisperse systems.
The comparison with numerically measured relaxation times
confirms the conclusions drawn from the study 
of the LJ potential that the Schweizer-Saltzman expression of the 
free-energy barrier considerably underestimates the slowing down
of the dynamics. This is more easily seen in the bottom panel which 
shows that numerical data have a much larger ``kinetic 
fragility'' (they are well  described\cite{berthier_witten} 
by $\tau \sim \exp(Z^\alpha)$ with $\alpha \approx 6$) than 
the theoretical ones, as also found for the LJ/WCA systems.
We note that the use of integral equations as input for the theory 
has two conflicting effects: on the one hand, integral equations 
tend to overestimate the structure of the fluid (which should
yield larger barriers), but on the other hand they are not sensitive
to the proximity of the jamming transition, a transition which is not captured
by simple liquid-state approaches (close to jamming this would
imply that barriers are underestimated). 

Finally, as for the LJ and WCA models in the main text, 
we use the numerically measured structure factor $S_{AA}(q)$ 
for the main component (larger particles) and  the 
effective one-component approximation for the free energy barrier. 
Although the approximation is expected to be less accurate in this 
50:50 mixture than for the 80:20 LJ/WCA systems, we find a trend
very similar to the one reported in the main text, namely
that integral equations seem to largely overestimate the evolution 
of the structure and that direct use of the ``exact'' structural 
information yields theoretical predictions in even stronger disagreement
with the simulations.

\begin{figure}[b]
\psfig{file=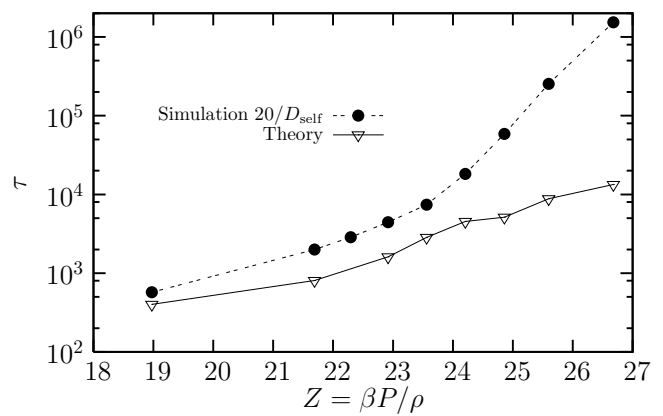,width=8.5cm}
\caption{\label{figure_4d}
Comparison of the dynamics of the four-dimensional monodisperse hard-sphere fluid obtained directly from simulations and obtained by using the 
Schweizer-Saltzman theoretical approach with the `exact' (numerically
determined) structure factor as an input. As in three dimensions, 
we find that using the exact structure produces free-energy barriers 
that are too small, and the observed glassy dynamics is 
poorly reproduced by the theory.}
\end{figure}

This short study of the hard-sphere systems therefore validates 
our implementation of the 
Schweizer-Saltzman approach for the binary LJ and WCA liquids
in the main text, and, as a way of consequence, the conclusions 
we have drawn from 
Fig.~\ref{figure_schweizer1}. \\

\section*{Addendum to the appendix}

If the assumptions we made are correct,
the last conclusions of the above appendix suggest that the agreement reported 
by Schweizer and Saltzman between their theoretical results and 
hard-sphere data is largely due a cancellation of errors between
a dynamical theory which underestimates barriers, and the use 
of integral equations which promotes them. As a further check
of our assumptions, we have studied the dynamics of a four-dimensional hard-sphere 
system. Indeed, in $4$ dimensions, a monodisperse system can be used with no 
crystallization issues~\cite{charbonneau}. In this case 
we can directly measure the `exact' structure factor
in the simulations and use the Schweizer-Saltzman theory to predict the dynamics, 
which can be compared to numerical measurements. In this comparison, 
the free-energy barriers are then estimated without approximations (whether due to polydispersity or to the computation of the structure factor)
and no assumption has to be made to implement the theory. 
We use the data obtained by Charbonneau and coworkers in 
Ref.~[\onlinecite{charbonneau}] 
to obtain the results shown in Fig.~\ref{figure_4d}. 
The results show the same features as in three dimensions, which therefore confirms the validity of 
our previous conclusions.

\end{document}